\documentclass[11pt]{article}
\usepackage{amsmath,euscript,amssymb}
\usepackage[dvips]{graphicx}
\def\v1{\vspace{1cm}}
\def\be{\begin{equation}}
\def\ee{\end{equation}}
\def\bc{\begin{center}}
\def\ec{\end{center}}

\def\vh{\varphi}

\newcommand{\bea}{\begin{eqnarray}}
\newcommand{\eea}{\end{eqnarray}}

\begin{document}
\title{QUANTUM COSMOLOGICAL ORIGIN OF UNIVERSES}
\author{ V.N. Pervushin, V.A. Zinchuk \\[0.3cm]
{\normalsize\it Joint Institute for Nuclear Research},\\
 {\normalsize\it 141980, Dubna, Russia.} }

\date{\empty}
\maketitle
\medskip

\begin{abstract}
 {\noindent
 A direct pathway from Hilbert's  ``Foundation
 of Physics'' to Quantum Gravity  is established through Dirac's Hamiltonian
 reduction of  General
 Relativity and Bogoliubov's transformation by analogy with a similar
 pathway passed by QFT in 20 th century.
  The cosmological scale factor
    appears on this pathway  as a zero mode of the momentum constraints
     treated as a global  excitation of the
   Landau superfluid liquid type. This approach would be considered
   as the  foundation of
    the well--known  Lifshitz cosmological perturbation theory,
     if it did not contain the double counting of the scale factor
     as an obstruction to the Dirac Hamiltonian method.
  After avoiding  this ``double counting''
   the Hamiltonian cosmological perturbation theory
 does not contain the time derivatives of gravitational potentials that
 are responsible for the CMB ``primordial power spectrum'' in
 the inflationary model.

  The Hilbert -- Dirac -- Bogoliubov Quantum Gravity
 gives us
 another  possibility  to explain this ``spectrum'' and
  other topical problems of cosmology by
    the cosmological creation of both universes and particles
  from  Bogoliubov's vacuum.
   We listed the set of theoretical and observational
 arguments in favor of that the CMB radiation can be
 a final product of  primordial vector W-, Z- bosons cosmologically created
 from the vacuum when  their Compton length coincides with the universe
 horizon.
 The equations describing longitudinal
 vector bosons
 in SM, in this case, are close to  the equations of the inflationary model
 used for
 description of the ``power primordial spectrum'' of the CMB radiation.
}
\end{abstract}

\vspace{-14cm}

{\bf To the 90th anniversary of GR and the 100th anniversary of
SR}

\vspace{13cm}

\vspace{3cm}

{\small \bf

~~Invited talk at the XXXIX PNPI Winter School on Nuclear Particle
Physics

\vspace{.1cm}

\begin{center}


 and XI St.Petersburg School on Theoretical Physics

\vspace{.3cm}

(St.Petersburg, Repino, February 14 - 20, 2005)

\vspace{.3cm}

{\bf +http://hepd.pnpi.spb.ru/WinterSchool/}

\end{center}
}
\newpage

\tableofcontents


\section{Introduction}

 It is now accepted that quantum General Relativity does  not
 exist. In the present paper, we give a set of theoretical and
 observational arguments in  favor of the opposite opinion:
 General Relativity (GR) \cite{H,einsh} has a consistent interpretation
 only in the form of quantum theory of the type of the microscopic
 theory of superfluidity \cite{Lon,Lan,B}.

 Our hopes for an opportunity to construct a
 realistic quantum theory for GR
 are based, on the one hand, on the existence of
 the Hilbert geometric formulation \cite{H,pp,bpp}  of Special
 Relativity  (SR) \cite{poi,ein} considered  as the simplest model of GR
 and, on the other hand, on  the contemporary quantum field theory
 (QFT) \cite{Schweber,Logunov} based on the dynamic version of SR
  \cite{poi,ein} that appears after resolution of constraints
 \cite{pp,bpp}.

 The geometric formulation of SR is an action of the type of
 the action in GR, which Hilbert has reported on 20th November 1915
 in his talk in G\"ottingen mathematical
 society  \cite{H}.

 We shall try here to reconstruct a direct pathway from
 geometry of a relativistic
 particle in SR to the  causal operator quantization of  fields
 of these particles and their quantum creation from vacuum in order to
  formulate a similar direct way from geometry of GR \cite{H}
 to the  causal operator quantization of universes
 and to their quantum creation from vacuum considered as
  a state with the minimal ``energy''. This formulation includes
the Wheeler -- DeWitt definition of ``field space of events''
\cite{WDW};
 the separation of gauge transformations from ones of the
 frames of references using the Fock simplex of reference \cite{fock29}; the choice
 of the Dirac specific frame of reference \cite{dir};
 resolving the energy constraint in the class of functions of the
 gauge transformations established by Zel'manov \cite{vlad}, where
 the cosmological scale factor appears as a  zero
 mode of the momentum constraints \cite{pp,bpp};
  the calculation of values of the geometric action and interval
 onto the resolutions of the energy constraint \cite{Dir}
 in order to get the dynamic
 ``reduced'' action  in terms of gauge-invariant
 variables and to define the notions ``energy``,
 ``time``, ``particle'' and ``universe'',
  ``number'' of ``particles'' and ``universes'' by
    the low-energy expansion of this ``reduced'' action following
  Einstein's correspondence principle with nonrelativistic
  theory \cite{poi,ein}.

In the present paper the cosmological evolution
\cite{Lif,kodama,bard} is treated as independent
 ``superfluid'' dynamics  \cite{Lan}.
One of attributes of the phenomenon of superfluidity
 is the application of  Bogoliubov's transformations \cite{B,ps1}
   to obtain
 a set of integrals of motions and calculate the distribution
 functions of cosmological creation of both ``universes'' and ``particles''
 from ``vacuum''. All these attributes:
  London's unique wave function \cite{Lon}, Landau's independent
 ``superfluid'' dynamics \cite{Lan}, and   Bogoliubov's transformations \cite{B}
  are accompanied by a set of physical consequences that can be
  understood as only pure quantum effects

 We shall show how this ``superfluid'' dynamics of GR gives us
 possible solutions of the topical astrophysical problems,
 including horizon, homogeneity, cosmological singularity, arrow of
 time,  Dark Matter, and Dark Energy.

 In Section 2,  the direct way from Hilbert's geometric formulation
 of SR to QFT is established. Section 3 is devoted to a similar
 Hilbert's foundation of quantum cosmology.
 In Section 4, GR is considered as a microscopic theory of
 superfluidity.

\section{Hilbert's Foundation of  Quantum Field Theory}

\subsection{Hilbert's version of Special Relativity}

  The Hilbert geometric formulation  of a relativistic particle
  \cite{H,pp,bpp}
  is based on  the action:
 \be\label{2} S^{\mbox{\tiny SR}}_{1915}
 =-\frac{m}{2}\int\limits_{\tau_1}^{\tau_2}d\tau
 \left[\frac{(\dot X_\alpha)^2}{e(\tau)}+e(\tau)
 \right],
 \ee
 and an {\it``geometric interval''}
  \be\label{3}
 ds=e(\tau)d\tau~~~~~\longmapsto~~~~~s(\tau)=
 \int\limits_{0}^{\tau}d\overline{\tau}e(\overline{\tau}),
 \ee
 where $\tau$ is the {\it``coordinate evolution parameter''}
 given in a one-dimensional Riemannian manifold with
 a single component of the metrics $e(\tau)$ and
 the variables $X_\alpha$ form the Minkowskian {\it``space of
 events''}, where $\left(X_\alpha\right)^2=X_0^2-X_i^2$.

 The action (\ref{2}) and interval (\ref{3}) are
 invariant with respect to reparametrizations of
 the {\it``coordinate evolution parameter''} (treated as {\it``gauge transformations''})
 \be\label{1}
 \tau ~~\longrightarrow~~~\widetilde{\tau}
 =\widetilde{\tau}(\tau);~
 \ee
 therefore, the theory given by (\ref{2}) and  (\ref{3})
 can be considered as the simplest model of GR.

 A single component of the metrics $e(\tau)$ (known as a
 {\it``lapse-function''})
  plays the role
 of the Lagrange multiplier in the Hamiltonian form of
 the action (\ref{2}):
 \be \label{4}
 S^{\mbox{\tiny SR}}_{1915}=\int\limits_{\tau_1}^{\tau_2}d\tau
 \left[-P_\alpha\partial_\tau X^\alpha+
 \frac{e(\tau)}{2m}\left(P_\alpha^2-m^2\right)\right].
 \ee
 Varying action  (\ref{4}) over lapse-function $e(\tau)$  defines the
 {\it``constraint''}:
 \be\label{co}
 (P_\alpha)^2-m^2=0.
 \ee
 Varying action  (\ref{4})  over
 dynamical variables $(P_{\alpha}, X_{\alpha})$
 gives the equations of motion:
 \be\label{eq}
 P_\alpha=m\frac{dX_\alpha}{ds},~~\frac{dP_\alpha}{ds}=0,
 \ee
 taking into
 consideration $ds=e(\tau)d\tau$.
 Solutions of equations (\ref{eq}) in terms of gauge-invariant
  {\it``geometric interval''} (\ref{3}) take the form
 \be  \label{5}
 X_\alpha(s)=X_\alpha(0)+
 \frac{P_\alpha(0)}{m}s.
 \ee

\subsection{Dirac's Hamiltonian reduction\label{dr}}

  The physical meaning of this solution is revealed  in
  a  specific {\it``frame of reference''}. In particular,
   solutions of energy constraint (\ref{co}) with respect to
  a temporal component
  $P_{0}$ of momentum $P_{\alpha}$
 \be  \label{6}
 {P_0}_{\pm}=\pm \sqrt{m^2+P_i^2}
 \ee
   are considered as the {\it``reduced
  Hamiltonian''} in the {\it``reduced phase space''} $\left\{X_i,P_j\right\}$
  that becomes the energy $E(P)=\sqrt{m^2+P_i^2}$ onto a trajectory
  \cite{poi,ein}. The time component of  solution
 (\ref{5})
 \be  \label{7}
 s=\frac{m}{{P_0}_{\pm}}[X_0(s)-X_0(0)] \ee
 shows us that the {\it``time-like variable''} $X_0$ is identified with the
 time measured in the rest frame of reference, whereas an interval $s$
 is the time measured in the comoving frame.

 The dynamic version of SR  \cite{poi,ein}
 can be obtained as values of the geometric action (\ref{4})
 onto solutions of the constraint (\ref{6})
 \be \label{8} S^{\mbox{\tiny SR}}_{1915}
|_{P_0={P_0}_{\pm}}=S^{\mbox{\tiny SR}}_{1905}
=\int\limits_{X_{0I}}^{X_0}d\overline{X}_0
 \left[P_i\frac{d{X}_i}{d\overline{X}_0}-{P_0}_{\pm}\right].
 \ee
  Just the values of the {\it`` geometric interval''} (\ref{7})
  and action (\ref{8}) onto  resolutions (\ref{6})
  of constraint (\ref{co}) in
  the specific frame of reference will be called
 the {\it ``Hamiltonian reduction''} of Hilbert's geometric
 formulation of SR given by Eqs. (\ref{2}) and  (\ref{3}) (see \cite{pp,Dir}).

\subsection{Dynamic version of Special Relativity of 1905}

 The {\it ``Hamiltonian reduction''} leads to action (\ref{8}) of
 the dynamic theory of a relativistic particle of ``1905''
 \cite{poi,ein} that
  establishes a correspondence with the classical
 mechanic action by the low-energy decomposition
  \be \label{9}
 E(P)=\sqrt{m^2+P_i^2}=m+\frac{P_i^2}{2m}+...
 \ee
 It gives us the very important  concept of particle ``energy'' $E(0)=mc^2$.
We can see that
 relativistic  relation (\ref{7})
between the ``time as the variable'' and the ``time as the
interval''
 appears in the geometric version  of ``1915'' \cite{H}
 as a consequence of the variational equations  (\ref{6}), whereas
 in the dynamic
 version of  ``1905'' \cite{poi,ein} the same relativistic
 relation in the form of a kinematic Lorenz
 relativistic transformation is supplemented to variational equations following
 from the dynamic action (\ref{8}).

\subsection{Quantum geometry of a relativistic particle}

 The next step forward to QFT is the primary quantization of
 particle variables: $i[\hat P_\mu,X_\nu]=\delta_{\mu\nu}$, that
 leads to
 the quantum version of the energy constraint (\ref{co})
 $[\Box +m^2]\psi(X_0,X_i)=0$ known  as the Klein -- Gordon equation
 of the wave function.
  The general solution of this
 equation
 \be\label{kg}
 \partial^2_0\Psi_p+E_p^2\Psi_p=0
 \ee
 for a single p-Fourier harmonics
 $\Psi_p(X_0)=\int d^3X \exp{iP_jX^j}\psi(X_0,X_i)$
 takes the form of the sum of two terms
 \be\nonumber
 \Psi_p=\frac{1}{\sqrt{2E_p}}\{a_{p}^{(+)}(X_0)+a_p^{(-)}(X_0)\},
 \ee
 where $a_{p}^{(+)}(X_0),a_p^{(-)}(X_0)$ are  solutions of the
 equations
\be \label{1pa} (i\partial_0+E_p)a_{p}^{(+)}=0,~~~~~~~~
(i\partial_0-E_p)a_p^{(-)}=0.
  \ee
 They are treated as the Shr\"odinger equations of the
 dynamic
 theory (\ref{8}) for the case of positive and negative
 particle ``energies'' (\ref{6}) revealed
 by resolving energy constraint (\ref{co}).



  QFT is formulated as the secondary quantization of a relativistic
 particle $[a_p^{(-)},a_{p}^{(+)}]=1$ \cite{Logunov}. In order to remove
 the negative ``energy'' $-E_p$ and to provide the quantum system with
 stability, the field $a_{p}^{(+)}$ is considered as the operator of
 creation of a particle and $a_{p}^{(-)}$  as the operator of
 annihilation of a particle, both with positive ``energy''.
 The initial datum $X_{I(0)}$ is treated as a point of this
 creation or annihilation. This interpretation means
 postulating vacuum as a state with minimal ``energy'' $a_p^{(-)} |0\rangle =0$,
 \label{pos} and it restricts the motion of a particle in
 the space of events, so that a particle with $P_{0+}$ moves
 forward and with $P_{0-}$ backward.
 \be \label{b12sr}
 P_{0+}~~~~ \to ~~~~X_{I{(0)}}\leq {X_{(0)}};~~~~~~~~~~~~~~~~~~
 ~~~P_{0-}~~~~ \to ~~~~X_{I{(0)}}\geq {X_{(0)}}.
  \ee
  As a result of such a restriction the interval (\ref{7})
  becomes
 \be  \label{7+}
 s_{(P_{0+})}=\frac{m}{{E_p}}[X_0(s)-X_0(0)];~~~~~~X_{I{(0)}}\leq {X_{(0)}},
 \ee
 \be  \label{7-}
 s_{(P_{0-})}=\frac{m}{{E_p}}[X_0(0)-X_0(s)];~~~~~~X_{I{(0)}}\geq {X_{(0)}}.
 \ee
 One can see that  in both cases the geometric interval is
 positive. In other words, the
 stability of quantum theory and the vacuum postulate as its consequence
 lead to the absolute reference point  of this interval $s=0$
 and its positive arrow.
 The last  means  violation of the symmetry of classical theory
 with respect to the transformation $s \to $ $-s$. Recall that the
 violation of the symmetry of classical theory by their
 quantization is called the quantum anomaly \cite{riv,ilieva,gip}.
 The quantum anomaly as the consequence of the vacuum postulate
 was firstly discovered by Jordan \cite{j} and then
 rediscovered by a lot of authors (see \cite{riv}).

 \subsection{Creation of particles}

  Creation of particles is described by QFT obtained by
   quantization of  classical fields with masses depending on time $m=m(X_0)$.
  Classical field equation (\ref{kg}) can be got by varying the action
  \be\label{ft1}
 S_{\rm p}=\int dX_0 \left\{P_p\partial_0\Psi_{p}-H_{\rm
 p}\right\},
 \ee
  where $H_{\rm p}=\frac{1}{2}\left[P_p^2+E_p^2(X_0)\Psi_p^2\right]$ is the
  field Hamiltonian,
   here we kept only one p-harmonics.

 \noindent The holomorphic representation of the fields \cite{ps1,skok}
 \be\label{g1}
 \Psi_p=\frac{1}{\sqrt{2E_p(X_0)}}\left\{a_{p}^{(+)}(X_0)+a_p^{(-)}(X_0)\right\},
 \ee
 \be\label{g2}
 P_p=i\sqrt{\frac{E_p(X_0)}{2}}\left\{a_{p}^{(+)}(X_0)-a_p^{(-)}(X_0)\right\}.
 \ee
 allows us
  to express the field Hamiltonian in action  (\ref{ft1})
 in terms of observable quantities --- the one-particle energy $E_p(X_0)$
 and  ``number'' of particles
 $N_p(X_0)=[a_{p}^+a_p^-]$:
 \be\label{ufh1z1}
 H_{\rm p}=\frac{1}{2}\left[P_p^2+E_p^2(X_0)\Psi_p^2\right]=
 E(X_0)\left[N_p(X_0)+\frac{1}{2}\right].
 \ee
   While the canonical structure
 $P_p \partial_0 \Psi$ in (\ref{ft1}) takes the form:
 \be\nonumber
   P_p \partial_0 \Psi_p=
  \left[\frac{i}{2}(a_p^+\partial_0 a_p^--a_p^+
 \partial_0 a^-)-
 \frac{i}{2}(a_p^+a_p^+- a_p^-a^-_p)\frac{\partial_0 E(X_0)}{2E(X_0)}\right].
 \ee
 The one-particle energy and the number of particle are not
 conserved. In order to find a set of  integrals of
 motion, we can use the Bogoliubov transformations
  \be\label{bogo}
  a_p^+=\alpha b_p^+\!+\!\beta^*b_p^-
  ~~~~~~~~~~~~~(\alpha=e^{i\theta}\cosh r,~~~
  ~\beta=e^{-i\theta}\sinh{r}),
  \ee
 so that  the equations of $b_p^+,b_p^-$ become diagonal
\be\label{bogo1}
 (i\partial_0+E_B)b_p^+=0,~~~~~~~~
 (i\partial_0-E_B)b_p^-=0,
 \ee
 and the conserved vacuum is defined by the postulate:
  \be\label{bogo2}
   b_p^-|0>_{\rm p}=0.
  \ee
 The  corresponding Bogoliubov equations of
 diagonalization expressed in terms  of the distribution function
 of {\it``particles''} $N_{\rm p}(X_0)$ and the rotation function
 $R_{\rm p}(X_0)$
 \be\nonumber
 N_{\rm p}(X_0)=|\beta|^2\equiv {}_{\rm p}<a_p^+a_p^- >_{\rm
 p}\equiv\sinh{r}^2,~~~~~~~~
 R_{\rm
 p}(X_0)=i(\alpha\beta^*-\alpha^*\beta)\equiv-\sin(2\theta)\sinh{2r}
 \ee
 take the form \cite{ps1}
 \be\label{1p}
 \left\{\begin{aligned}
 \frac{dN_{\rm p}}{dX_0}&=&\frac{\partial_0 E(X_0)}{2E(X_0)}
 \sqrt{4N_{\rm p}(N_{\rm p}+1)-R_{\rm p}^2},  \\
 \frac{dR_{\rm p}}{dX_0}&=&-{2E(X_0)}
 \sqrt{4N_{\rm p}(N_{\rm p}+1)-R_{\rm p}^2},
 \end{aligned}\right.
 \ee
 \be \label{6pa}
 E_B(X_0)=\frac{E_p(X_0)-\partial_0\theta}{\cosh 2r}.
 \ee
 These equations supplemented by the quantum geometric interval
 (\ref{7+}) and (\ref{7-}) are the complete set of equations  for description
 of the phenomenon of particle creation.

 Thus, the direct way from
 Hilbert's geometric formulation of any relativistic theory to the
 corresponding
 ``quantum field theory''
 goes through  Dirac's Hamiltonian reduction and Bogoliubov's
 transformations. As a result, we have
 the description of creation of a relativistic particle in the space of events
 at the absolute reference point
 of geometric interval $s$ of this particle. The physical meaning of this
 interval is revealed in the Quantum Cosmology considered below.

\section{Quantum Cosmology}

\subsection{Hilbert's ``Foundation of Physics'' of 1915 \cite{H}}

  Einstein's GR \cite{einsh} is based on the dynamic action
  \be\label{h1}
 S_{GR}=\int d^4x\sqrt{-g}
 \left[-\frac{\vh_0^2}{6}~~R(g)+{\cal L}_{\rm
 matter}\right],~~~~~\vh_0^2=\frac{3}{8\pi}M^2_{Planck}
 \ee
 proposed
  by Hilbert in his talk  on 20th November 1915
  in G\"ottingen mathematical
 society  \cite{H}; this action is given in a Riemannian
 space-time manifold with  {\it``geometric interval''}
 \be\label{h2}
 ds=g_{\mu\nu}dx^{\mu}dx^{\nu}.
 \ee
  Both the action and interval are invariant with respect
  to general coordinate transformations
  \be\label{gct}
 x^{\mu}~~~~~\longrightarrow~~~~~\tilde{x}^{\mu}=
 \tilde{x}^{\mu}(x^0,x^1,x^2,x^3).
 \ee
 GR (\ref{h1}), (\ref{h2}) is similar to the geometric version of SR considered
 in the previous section. Therefore, we can repeat the pathway of
 SR to QFT of particles considered in Section 2.

\subsection{Foundation of Quantum Cosmology of 2005}

 In order to demonstrate to a reader the direct pathway from
  Hilbert's ``Foundation of Physics''  to QFT of universes
 through Dirac's Hamiltonian reduction \cite{pp}, we consider GR
 in the homogeneous approximation of the
 interval
 $$
 ds^2\simeq ds^2_{\rm WDW}=a^2(x^0)[(N_0(x^0)dx^0)^2 -(dx^idx^i)].
 $$
 where Hilbert's action  (\ref{h1}) and interval  (\ref{h2}) take the form
\be\label{H1}
 S_{\rm cosmic-1915}=V_0\int
 dx^0N_0\left[-\left(\frac{d\varphi}{N_0dx^0}
\right)^2-\rho_0(\varphi)\right],  
 \ee
\be\label{conf}
 d\eta=N_0(x^0)dx^0~~~\longrightarrow~~
 ~~\eta=\int\limits_{0}^{x^0} d\overline{x}^0 N_0(\overline{x}^0),
 \ee
 here $a(x^0)$ is the  cosmological scale
 factor, $\vh(x^0)=\vh_0a(x^0)$,
 $\rho_0(\vh)$ is the energy density of the matter in a universe,
 $V_0$ is a spatial volume,
  and $N_0$ is the lapse function.
 This homogeneous approximation keeps the symmetry of the action
 and the  interval with respect to
  reparametrizations of the  coordinate evolution
 parameter $x^0~\to~\overline{x}^0=\overline{x}^0(x^0)$.
 Recall that similar transformations in the case of SR play
 the role of gauge transformations; gauge invariance determines
 observable quantities of the type of energy, time-like variable and number of
 particles.

 In the WDW cosmology this gauge symmetry
 means that the scale factor $\vh$
 is the {\it``time-like variable''} in the {\it``field space of events''}
 introduced
 by Wheeler and DeWitt
 in \cite{WDW}, and the canonical momentum $P_\vh$ is the corresponding
 Hamiltonian that becomes the energy $E_\vh$ into equations of motion.

 The direct pathway from Hilbert's SR to QFT of particles
 considered in Section 2 shows us a similar direct pathway
  from  Hilbert's ``Foundation of Cosmology'' (\ref{H1}), (\ref{conf})
 to ``QFT'' of universes.
 This pathway includes:

1) Hamiltonian approach: \be\label{ham}
 S_{\rm cosmic-1915}=\int
 dx^0\left[-P_\vh\partial_0
 \vh-\frac{N_0}{4V_0}\left(-P^2_\vh+E^2_\varphi\right)\right],
 \ee
 where
 \be\label{ev1}
 E_\varphi=2V_0\sqrt{\rho_0(\vh)}
 \ee
 is treated as the {\it''energy of a universe''},

2) constraining: $P^2_{\varphi}-E^2_\varphi=0$,

3) primary quantization:
$[\hat{P}^2_{\varphi}-E^2_\varphi]\Psi=0$, here
$\hat{P}_{\varphi}=-id/d\varphi$,

4) secondary quantization: $\Psi=([A^++A^-]/\sqrt{2E_\vh})$,

5)  Bogoliubov's transformation:
 $ A^+=\alpha
 B^+\!+\!\beta^*B^-$,

6) postulate of Bogoliubov's vacuum : $B^-|0>_{U}=0$, and

7) cosmological creation of the
 {\it ``universes''} from the Bogoliubov vacuum.

 Let us carry out this programme  in detail.

\subsection{{``Hamiltonian reduction''}}

 In the cosmological model (\ref{ham}), there are two independent
  equations: the one  of the lapse function
 $\delta S_{\rm cosmic-1915}/\delta N_0=0$:
 \be\label{1conn}
 P^2_\vh=E_\vh^2,
 \ee
 treated as the  energy
 constraint, and the equation of momentum
 $\delta S_{\rm cosmic-1915}/\delta P_\vh=0$
 \be\label{3conn}
 {P}_\vh= 2V_0\vh',
 \ee
  where $\vh'=\frac{d\vh}{d\eta}$.
 The constraint (\ref{1conn}) has two solutions
 \be\label{2conn}
 {P^{\pm}_\vh}=\pm E_\vh=\pm 2V_0\sqrt{\rho(\vh)},
 \ee
 where $E_\vh$, given by Eq. (\ref{ev1}), is identified with
 the ``one-universe energy''. The substitution of these solutions
 into action (\ref{ham}) and interval  (\ref{conf}) gives us
 their values
\be\label{ham-2}
 {S_{\rm cosmic-1915}}_{|P_\vh=P^{\pm}_\vh}=S^{\pm}_{\rm cosmic-1905}=\mp
2V_0\int\limits_{\vh}^{\vh_0}
 d\widetilde{\vh}\sqrt{\rho_0(\widetilde{\vh})},
 \ee
 and
\be\label{4conn}
 \eta(\vh|\vh_0)=2V_0
 \int\limits_{\vh}^{\vh_0}\frac{d\widetilde{\vh}}
 {{P^{\pm}_\vh}}=\pm
 \int\limits_{\vh}^{\vh_0}\frac{d\widetilde{\vh}}
 {\sqrt{\rho_0(\widetilde{\vh})}}\equiv\pm (\eta_0-r),
 \ee
 respectively. We called these values
 the {\it``Hamiltonian reduction''} of the geometric system
 (\ref{ham}), (\ref{conf}).
 Eq. (\ref{4conn}) is treated,
  in the observational cosmology \cite{Narlikar,039,039a,Danilo},
  as the conformal version of the
  Hubble law. This law
   describes the relation between the redshift $z+1=\vh_0/\vh(\eta)$
  of spectral lines of photons (emitted by atoms on a cosmic object
  at the conformal time $\eta(\vh|\vh_0)=\eta$) and
  the coordinate distance $r=\eta_0-\eta$ of this object, where
  $\eta_0$ is the present-day moment. Thus, we see that
  WDW cosmology coincides with the Friedmann one
  $\vh{'}{}^2=\rho_0(\vh)$. Our task is to consider the status of this
 Friedmann cosmology in QFT of universes obtained by
 the first and the secondary quantization of the constraint
 (\ref{1conn}).

 \subsection{QFT of universes}
%
%

 After the primary  quantization of the cosmological scale factor $\vh$:
 $i[P_\vh,\vh]=1$ the energy constraint (\ref{1conn}) transforms
 to the WDW equation
\be\label{wdw}
 \partial^2_\vh\Psi+E_\vh^2\Psi=0.
 \ee
 This equation can be obtained in the corresponding classical
 WDW field theory for universes of the type of the Klein -- Gordon
 one:
 \be\label{uf}
 S_{\rm U}=\int d\vh \frac{1}{2}
 \left[(\partial_\vh\Psi)^2-E_\vh^2\Psi^2\right]\equiv \int d\vh
  L_{\rm U}.
 \ee
 Introducing the canonical momentum
$P_\Psi=\partial L_{\rm U}/\partial(\partial_\vh\Psi)$, one can
obtain the Hamiltonian form of this theory
 \be\label{ufh}
 S_{\rm U}=\int d\vh \left\{P_\Psi\partial_\vh\Psi-H_{\rm
 U}\right\},
 \ee
 where
\be\label{ufh1}
 H_{\rm U}=\frac{1}{2}\left[P_\Psi^2+E_\vh^2\Psi^2\right].
 \ee
 is the Hamiltonian. The concept of the one-universe ``energy'' $E_\vh$
 gives us the opportunity to present this Hamiltonian $H_{\rm U}$
 in the standard forms of the product of this ``energy''  $E_\vh$ and
 the ``number'' of universes
 \be\label{AA}
 N_U=A^+A^-,
 \ee
 \be\label{ufh1z12}
 H_{\rm U}=E_\vh\frac{1}{2}\left[A^+A^-+A^-A^+\right]=E_\vh[N_U+\frac{1}{2}]
 \ee
 by  means of the transition to the holomorphic variables
 \be\label{g}
 \Psi=\frac{1}{\sqrt{2E_\vh}}\{A^{(+)}+A^{(-)}\},~~~~~~~~
 P_\Psi=i\sqrt{\frac{E_\vh}{2}}\{A^{(+)}-A^{(-)}\}.
 \ee
 The  dependence of $E_\vh$  on $\vh$ leads to the additional term
 in the action expressed in terms the holomorphic variables
 \be\label{und3}
   P_\Psi \partial_\vh \Psi=
  \left[\frac{i}{2}(A^+\partial_\vh A^--A^+
 \partial_\vh A^-)-
 \frac{i}{2}(A^+A^+- AA)\triangle (\vh)\right],
\ee
 where
 \be\label{tri}
 \triangle(\vh)=\frac{\partial_\vh E_\vh}{2E_\vh}.
 \ee
  The last term in (\ref{und3}) is responsible for
  the cosmological creation of  ``universes'' from ``vacuum''.

 \subsection{Creation of universes}

 In order to define stationary physical states, including
 a ``vacuum'', and a set of integrals of motion, one usually uses
 the Bogoliubov transformations \cite{B,ps1} of the holomorphic
 variables of universes
 $(A^+,A^-)$:
 \be \label{u17} A^+=\alpha
 B^+\!+\!\beta^*B^-,~~~~~~~\;\;A^-=\alpha^*
 B^-\!+\!\beta A^+~~~~~~~~~~~(|\alpha|^2-|\beta|^2=1),
  \ee
  so that the classical equations of the field theory  in terms of universes
 \be \label{1un} (i\partial_\vh+E_\vh)A^+=iA^-\triangle(\vh),~~~~~~~~
(i\partial_\vh-E_\vh)A^-=iA^+\triangle(\vh),
  \ee
  take a diagonal form in terms of  {\it``quasiuniverses''}
  $B^+,B^-$:
 \be \label{2un} (i\partial_\vh+E_B(\vh))B^+=0,~~~~~~~~
 (i\partial_\vh-E_B(\vh))B^-=0.
  \ee
  The diagonal form is possible, if the  Bogoliubov coefficients $\alpha,\beta$ in
  Eqs. (\ref{u17})
  satisfy to equations
 \be \label{3un} (i\partial_\vh+E_\vh)\alpha=i\beta\triangle(\vh),~~~~~~~~
(i\partial_\vh-E_\vh)\beta^*=i\alpha^*\triangle(\vh).
  \ee
 For the parametrization
 \be \label{4un} \alpha=e^{i\theta(\vh)}\cosh r(\vh),~~~~~~~~
 \beta^*=e^{i\theta(\vh)}\sinh{r}(\vh),
  \ee
  where $r(\vh),\theta(\vh)$ are the parameters of ``squeezing''
  and ``rotation'', respectively, Eqs. (\ref{3un}) become
 \be \label{5un}
 (i\partial_\vh\theta-E_\vh)\sinh 2r(\vh)=-\triangle(\vh)\cosh 2r(\vh)\sin
 2\theta(\vh),~~~~~~~~\partial_\vh r(\vh) =\triangle(\vh)\cos
 2\theta(\vh),
 \ee
  while ``energy'' of  {\it``quasiuniverses''} in Eqs. (\ref{2un})
  is defined by expression
   \be \label{6un}
  E_B(\vh)=\frac{E_\vh-\partial_\vh\theta(\vh)}{\cosh 2r(\vh)}.
  \ee
 Due to Eqs. (\ref{2un}) the ``number'' of {\it``quasiuniverses''}
  ${\cal N}_B=(B^+B^-)$ is conserved
  \be \frac{d{\cal N}_B}{d\vh}\equiv
  \frac{d(B^+B^-)}{d\vh}=0.
  \ee
  Therefore, we can introduce the ``vacuum''
 as a state without {\it``quasiuniverses''}:
 \be \label{sv}
 B^-|0>_{\rm U}=0.
 \ee
 A number of created {\it``universes''} from this Bogoliubov vacuum
 is equal to the expectation value of the operator
 of the  {\it``number of universes''} (\ref{AA}) over the Bogoliubov
 vacuum
 \be\label{usv1}
 N_{\rm U}(\vh)={}_{\rm U}<A^+A^-
  >_{\rm U}\equiv |\beta|^2=\sinh^2r(\vh),
 \ee
 where $\beta$ is the coefficient in the Bogoliubov transformation
 (\ref{u17}), and $N_{\rm U}(\vh)$ is called the ``distribution
 function''. Introducing the ``rotation function''
\be\label{usv2}
 R_{\rm U}(\vh)=i(\alpha\beta^*-\alpha^*\beta)
 \equiv {}_{\rm U}<P_\Psi\Psi >_{\rm U},
 \ee
 one can rewrite the Bogoliubov equations of the diagonalization (\ref{3un})
  in the form of (\ref{1p})
\be\label{usv3}
 \left\{\begin{aligned}
 \frac{dN_{\rm U}}{d\vh}&=\triangle(\vh)
 \sqrt{4N_{\rm U}(N_{\rm U}+1)-R_{\rm U}^2},  \\
 \frac{dR_{\rm U}}{d\vh}&=-{2E_\vh}
 \sqrt{4N_{\rm U}(N_{\rm U}+1)-R_{\rm U}^2}.
 \end{aligned}\right.
 \ee
  It is natural to propose that at the moment of creation of
  the universe $\vh(\eta=0)=\vh_I$ both these functions are equal to zero
   $N_{\rm U}(\vh=\vh_I)=R_{\rm U}(\vh=\vh_I)=0$.
  This moment of the conformal time (\ref{4conn})
  $\eta=0$
   is distinguished by the vacuum postulate (\ref{sv}) as the beginning
   of a universe.

\subsection{Quantum anomaly of conformal time}

 As we have seen in the case of a particle in Section \ref{pos},
 the postulate of a vacuum as a state with minimal ``energy''
 restricts the motion of a ``universe'' in
 the space of events, so that a ``universe'' with $P_{\vh+}$ moves
 forward and with $P_{\vh-}$ backward.
 \be \label{1b12sr}
 P_{\vh+}~~~~ \to ~~~~\vh_{I}\leq {\vh_{0}};~~~~~~~~~~~~~~~~~~
 ~~~P_{\vh-}~~~~ \to ~~~~\vh_{I}\geq {\vh_{0}}.
  \ee
  If we substitute this restriction into the interval (\ref{4conn})
 \be  \label{u7+}
 \eta_{(P_{vh+})}=\int\limits_{\vh_I}^{\vh_0}\frac{d\vh}{\sqrt{{\rho_0(\vh)}}}
 ;~~~~~~~~~~~~~~~~~~\vh_{I}\leq {\vh_{0}},
 \ee
 \be  \label{u7-}
 \eta_{(P_{\vh-})}=\int\limits_{\vh_0}^{\vh_I}\frac{d\vh}{\sqrt{\rho_0(\vh)}}
 ;~~~~~~~~~~~~~~~~~~\vh_{I}\geq {\vh_{0}},
 \ee
 one can see that the geometric interval in both cases is
 positive. In other words, the stability of quantum theory
 as the vacuum postulate leads to the absolute point of
 reference of this interval $s=0$  and its positive arrow.
   In QFT the initial datum $\vh_I$ is considered  as a point of
 creation or annihilation of universe.
 One can propose that the singular point $\vh=0$ belongs to
 antiuniverse. In this  case, a universe with a positive
 energy goes out of the singular point   $\vh =0$.

 In the model of rigid state $\rho=p$, where  $E_\vh=Q/\vh$
 Eqs. (\ref{usv3}) have an exact solution
\be\label{11cu}
 N_{\rm U}=\frac{1}{4Q^2-1}
 \sin^2\left[\sqrt{Q^2-\frac{1}{4}}~~\ln\frac{\vh}{\vh_I}\right]\not
 =0,
\ee
 where
 \be\label{cc}
 \vh=\vh_I\sqrt{1+2H_I\eta}
 \ee
  and
$\vh_I,H_I=\vh'_I/\vh_I=Q/(2V_0\vh_I^2)$ are the initial data.

 We see that there are results of the type of the arrow of
 time and absence of the cosmological singularity (\ref{u7+}),
 which can be understood only on the level of
  quantum theory \cite{riv,ilieva,gip}.

\section{General Relativity as a microscopic theory of superfluidity}

\subsection{``Foundation of Physics'' in terms of Fock's  simplex of reference}

     General Relativity (GR)  \cite{einsh,H}
    is given by two fundamental quantities:
    the  {\it``dynamic''} action (\ref{h1})
 \be\label{gr}
 S[\vh_0|F]=\int d^4x\sqrt{-g}\left[-\frac{\vh_0^2}{6}R(g)
 +{\cal L}_{(\rm M)}(\vh_0|g,f)\right],
 \ee
 where  $\varphi^2_0=\frac{3}{8\pi}M^2_{\rm Planck}$ is the Newton
constant,
 ${\cal L}_{(\rm M)}$  is the  Lagrangian of the matter field $f$, $F=(g,f)$,
     and
 {\it``geometric interval''}  (\ref{h2})
\be \label{ds}
 g_{\mu\nu}dx^\mu dx^\nu\equiv\omega_{(\alpha)}\omega_{(\alpha)}=
 \omega_{(0)}\omega_{(0)}-
 \omega_{(1)}\omega_{(1)}-\omega_{(2)}\omega_{(2)}-\omega_{(3)}\omega_{(3)},
 \ee
 where $\omega_{(\alpha)}$ linear differential forms introduced
 by Fock  \cite{fock29} as components of an orthogonal
 simplex of reference.

 Hilbert's
 {\it``Foundation of Physics''}
 in terms of Fock's  simplex  (\ref{gr}), (\ref{ds})
  contains  two principles of
 relativity: the {\it``geometric''} --- general coordinate transformations
\be \label{1zel}
 x^{\mu} \rightarrow  \tilde x^{\mu}=\tilde
 x^{\mu}(x^0,x^{1},x^{2},x^{3}),~~~~~~~~~
 \omega_{(\alpha)}(x^{\mu})~\to ~\omega_{(\alpha)}(\tilde x^{\mu})=
 \omega_{(\alpha)}(x^{\mu})
 \ee
 and the {\it``dynamic''} principle formulated as the Lorentz
 transformations of an orthogonal  simplex of reference
 \be \label{2zel}
{\omega}_{(\alpha)}~\to ~
\overline{\omega}_{(\alpha)}=L_{(\alpha)(\beta)}{\omega}_{(\beta)}.
\ee
  The latter are considered as transformations of a frame of reference.

  Fock's  separation of  the frame transformations  (\ref{2zel})
 from the gauge ones (\ref{1zel})
  \cite{fock29} allows us to consider GR and SR on equal footing.
Therefore, we shall try to reconstruct a direct pathway from
Hilbert's
 geometric formulation of GR to Quantum Gravity
through Dirac's Hamiltonian reduction and Bogoliubov
 transformations:

\vspace{4mm}

 \begin{tabular}{|p{15cm}|}
 \hline

 \vspace{2mm}
\label{gd} ~~~~~~~~~~~~~~~~~~~~\fbox{\rm
GR-1915}~~~~~~~~~~~~~~~~~~~~~~~~\fbox{\rm SR-1915}
 ~~~~~~~~~~~~
 $$
 ~~~~~~~~~~~~~~~~~~~~~~~ ~\Downarrow~~~~~~~~~~~~~~~~~~~~~~~~ ~~~~~~~~~~ \Downarrow
 ~~~~~~~~~~~~~~~~~~\Leftarrow~~~~~~~~\fbox{\rm reduction}~~~~~~~
 $$
 $$
\fbox{\rm GR-1905}~~~~~~~~~~~~~~~~~~~~~~~~\fbox{\rm
SR-1905}~~~~~~~~~~~~~~~~~~~~~~~~~~~
 $$
$$
 ~~~~~~~~~~~~~~~~~~~~~~~~~~~\Downarrow ~~~~~~~~~~~ ~~~~~~~~~~~~~~~ ~~~~~~~~~~~~\hfill \Downarrow
 ~~~~~~~~~~~~~~~~~~~~\Leftarrow~~~~~~\fbox{\rm
 quantization}~~~~~~~
$$
$$
~~~~~~~~~~~\fbox{\rm QFT~of~ universes}~~~~~~~~~
 ~~~\fbox{\rm QFT~of~particles}~~~~~~~~~~~~~~~~~~~~~~~~~~~~~~
 ~~~~~~~~
$$\\
\hline
\end{tabular}

\vspace{4mm}

 Let us consider this pathway that was almost passed by
 Fock, Dirac, and other physics.

\subsection{\label{adm1} The Dirac -- ADM frame}

  The Hamiltonian approach to GR is formulated
  in the frame of reference  given by  Fock's simplex of reference
  in terms of the Dirac variables \cite{dir}
\be \label{adm}
 \omega_{(0)}=\psi^6N_{\rm d}dx^0,~~~~~~~~~~~
 \omega_{(b)}=\psi^2 {\bf e}_{(b)i}(dx^i+N^i dx^0);
 \ee
 here triads ${\bf e_{(a)i}}$ form the spatial metrics with $\det |{\bf
 e}|=1$.

\subsection{\label{fock} Wheeler-DeWitt relativistic universe}
  A {\it``universe''}  is   a solution of the
 Einstein equations that describes a hypersurface in
 the {\it``field space of events''} identified with the set of all field
     variables $F=(g,f)$ \cite{WDW} in a specific frame of reference (\ref{adm}).
 There are two types of the variational equations:
 six equations of motion
\be\label{e}
 \frac{\delta S}{\delta\psi}=0, ~~~~~~ \frac{\delta S}{\delta{\bf e}_{(a) i}}=0
\ee and four constraints
 \be\label{c}
 \frac{\delta S}{\delta N_d}=0, ~~~~~~
 \frac{\delta S}{\delta N^k}\equiv {T^0_k}_t(\vh_0|F)=0.
\ee
 The general solution of these equations should be given in the
 class of functions of the gauge transformations in terms of
 the Dirac gauge-invariant observables in each frame of reference, so that
 the {\it``dynamic principle''} is formulated as independence of
 equations of motion (but not their solutions)
  on a choice of a frame of reference, and
 the {\it``geometric principle''} is treated in \cite{pp,Dir}
 as the gauge invariance of
 observables.

\subsection{\label{ze}  Zel'manov's class of functions of the
 gauge transformations}

  Zel'manov established in \cite{vlad} that a frame of reference determined by
   forms (\ref{adm}) is invariant with respect to
  transformations
 \be \label{zel}
 x^0 \rightarrow \tilde x^0=\tilde x^0(x^0);~~~~~
 x_{i} \rightarrow  \tilde x_{i}=\tilde x_{i}(x^0,x_{1},x_{2},x_{3})~,
 \ee
 \be \label{kine}
 \tilde N_d = N_d \frac{dx^0}{d\tilde x^0};~~~~\tilde N^k=N^i
 \frac{\partial \tilde x^k }{\partial x_i}\frac{dx^0}{d\tilde x^0} -
 \frac{\partial \tilde x^k }{\partial x_i}
 \frac{\partial x^i}{\partial \tilde x^0}~.
 \ee
 This group of transformations conserves
  a family  of hypersurfaces  $x^0=\rm{const.}$,
  and it calls the {\it``kinemetric''} subgroup of the group of
  general coordinate  transformations.
    The {\it``kinemetric''} subgroup contains
 reparametrizations of the coordinate evolution parameter.

 The reparametrization of the coordinate evolution parameter $(x^0)$
   means that this specific frame of reference
  (\ref{adm}) should be redefined by pointing out two {\it``Dirac observables''}:
{\it``time-like variable''} in the {\it``field space of events''}
and  the {``time''} as a {\it``geometric
 interval''} \cite{pp,bpp,ps1}.

\subsection{\label{li} Cosmological scale factor as zero mode of the momentum constraints}

 One of the main problems of the Hamiltonian approach to GR is to
 pick out the global variable which can play the role of
{\it``time-like variable''}.
 There is a lot of speculations on this subject
 \cite{pp,bpp,ps1,Y,prd}. In  \cite{pp,bpp,ps1} one proposed
 to  identify this
 ``internal evolution parameter''  with the  cosmological
 scale factor  $a(x_0)$ considered as a zero mode of the momentum
 constraints (\ref{c}) ${T^0_k}_t(F)=0$ \cite{pp,Dir,gip} given in
the class of functions of Zel'manov's gauge transformations
(\ref{zel}) that includes homogeneous functions $a(x^0)$.

Separation of real dynamical variables from nondynamical ones
 is the crucial step in extracting the relevant physical
    information from the gauge theories. The usual method for achieving this
purpose --- by imposing a gauge--fixing condition, might not be
always adequate to the dynamical content of the classical
equations of motion. Another possibility, offered by Dirac
\cite{Dir}, consists in introduction of gauge invariant dynamical
variables through an explicit solution of the
    Gauss equation. Such an explicit solution might contain some additional
    physical information which is implicitly lost by the gauge fixing.
    This can be seen even on the simplest example of the
    two--dimensional QED \cite{gip}, where the Gauss constraint
 $\partial_1 E(x^0,x^1)=0$ has a nontrivial homogeneous solution
 $E(x^0,x^1)=E_0(x^0)$ that determines the topological structure
 and energy spectrum of the theory; and this  solution is called
   {\it``zero mode''}. There is
 a similar nontrivial homogeneous solution of the constraint (\ref{c})
 ${T^0_k}_t(\vh_0|F) =0$ in GR.

  Constraints ${T^0_k}_t(\vh_0|F)=0$ are
 invariant  with respect to
the Lichnerowicz scale transformation \cite{L}:
 $F^{(n)}=a^n(x_0) \overline{F^{(n)}}$, where $(n)$ is the conformal weight
 of a field:
 ${T^0_k}_t(\vh|\overline{F})={T^0_k}_t(\vh_0|F)$, like the Gauss constraint
 in two--dimensional QED \cite{gip}
 $\partial_1 E(x^0,x^1)=0$ is invariant
 with respect to the  transformation
 $E(x^0,x^1)=E_0(x^0)+\overline{E}(x^0,x^1)$. Therefore, in the Hamiltonian
 approach to GR, the scale
 $a(x^0)$ is considered  as a  {\it``zero mode''} of the
 momentum constraints
 in the definite frame of
 reference.

Thus, a
 general solution of constraints ${T^0_k}_t(\vh_0|F)=0$ in the class
 of functions (\ref{zel}) can be written in the form of
 the Lichnerowicz scale transformation \cite{L}.
  In particular,
 the spatial metric determinant takes the form
 $\psi^2=a(x_0)\overline{\psi}^2$. The last equation can be treated
 as the decomposition of the logarithm of spatial metric
 determinant in the form of a sum of zero-Fourier harmonics
 $\left\langle{ \log{\psi^2}}\right\rangle \equiv \int d^3x
 \log{\psi^2}/V_0$ (where $V_0=\int d^3x  < \infty$ is a finite volume)
 and nonzero ones distinguished by the identity
 \be\label{non1}
 \int d^3x \log\overline{\psi}^2=\int d^3x \left[\log{\psi^2}
 -\left\langle{ \log{\psi^2}}\right\rangle\right]\equiv 0.
 \ee

\subsection{ Separation  of the zero mode by Lichnerowicz's  transformation}

 In order to find values of the action (\ref{gr}) into resolutions of
 constraints $g_{\mu\nu}=a^2(x^0)\overline{g}_{\mu\nu}$,
 we redefine a specific frame of reference (\ref{adm}) by
 the Lichnerowicz scale transformation $F^{(n)}=a^n(x_0)
 \overline{F^{(n)}}$.
 The Einstein -- Hilbert action (\ref{gr})
 after the scale transformation
  takes form
 \be\label{1gr}
 S[\vh_0|F]=S[\vh|\overline{F}]+
 \int dx^0\vh\partial_0\left[\frac{\partial_0\vh}{N_0}\right],
 \ee
 where
 ${N_0(x^0)}^{-1}={V^{-1}_{0}}\int_{V_0}d^3x{ \overline{N}^{-1}_d(x^0, x^i)}
 \equiv\left\langle{ \overline{N}^{-1}_d}\right\rangle
 $
 is the averaging of the  inverse  lapse
 function $\overline{N}_d$ over  spatial volume $V_0=\int d^3 x$,
 \be \label{sv11}
 S[\varphi|\overline{F}]= \int d^4x
 \left[{\bf K}[\vh| \overline{g}]-{\bf P}[\vh|\overline{g}]+
{\bf S}[\overline{g}]+
 {\cal L}_{(\rm M)}(\vh| \overline{F})\right]
 \ee
  is the action (\ref{gr}), where  $[\vh_0|F]$ is replaced by
 $[\varphi|\overline{F}]$; here
 $\vh(x^0)=\vh_0a(x^0)$ is the running scale of all masses  of the
 matter field,
 \be\label{k1}
 {\bf K}[\vh| \overline{g}]=\vh^2{N}_d\left[-{\vphantom{\int}}4
 {  v_{\psi}}^2+\frac{v^2_{(ab)}}{6}]\right],
 \ee
\be\label{p1}
 {\bf P}[\vh|\overline{g}]=\frac{{\vh^2}\psi^7}{{6}}
 \left[{}^{(3)}R({\bf e})\psi+8\triangle\psi\right],
 \ee
 \be\label{S1}{\bf
 S}[\vh|\overline{g}]=2\varphi^2\left[\partial_0v_{\psi}-
 \partial_l(N^l v_{\psi})\right]-
 \frac{\varphi^2}3
\partial_j[\overline{\psi}^2\partial^j (\overline{\psi}^6
 \overline{N}_d)];
 \ee
 are the kinetic, potential, and ``quasi--surface'' terms
 respectively,
 \be
 v_{(ab)}=\frac{1}{2}\left({\bf e}_{(a)i}v^i_{(b)}+{\bf
 e}_{(b)i}v^i_{(a)}\right),~~~~~~~v^i_{(a)}=\frac{1}{\overline{N_d}}\left[
 (\partial_0-N^l\partial_l){\bf e}^i_{(a)}+\partial_{(a)}N^i-
 \frac{{\bf e}^i_{(a)}}{3}\partial_lN^l\right]
 \ee
 are velocities of
  triads ${\bf e_{(a)i}}$, ${}^{(3)}R({\bf e})$ is the curvature of
  the triads,  and
\be\label{pi1}
 v_{\psi}=\frac{1}{\overline{N}_d}\left[
 (\partial_0-N^l\partial_l)\ln{\overline{\psi}}-\frac16\partial_lN^l\right]
 \ee
 is the trace of the ``second form''.

The last term in Eq. (\ref{1gr})
  determines the {\it``time as interval''} in Dirac's frame (\ref{adm})
 \be\label{ght} d\zeta=N_0(x^0)dx^0; ~~
 ~\zeta(x^0)=\int\limits_{}^{x^0}d\overline{x}^0N_0(\overline{x}^0).
 \ee

\subsection{ Avoiding  double counting of canonical momenta}

  Neglecting total time
 derivatives, we keep in Eq. (\ref{1gr}) only  the part of
Lagrangian describing the spatial metric determinant:
\be\label{sd}
 L_{\rm SD}=-\int d^3x{N}_d
 \left[{\vphantom{\int}}4\vh^2~
 {  (v_{\psi})}^2
 +4\vh~ v_\vh~  v_{\psi}+
 ({v_\vh)^2}{\vphantom{\int}}\right],
 \ee
 where $v_\vh=\partial_0\vh/N_d$,
 the first term in the Lagrangian arises from the kinetic part
${\bf K}[\vh|\overline{g}]$ in Eq. (\ref{sv11}),
 the second goes from the ``quasi-surface'' one (\ref{S1}),  and the
 third term goes from the second one in Eq. (\ref{1gr}).
 The canonical
 momentum of the scale factor can be obtained by variation of
 Lagrangian  (\ref{sd})
 with respect to  the time derivative of scale factor $\partial_0\varphi$:
$$
 P_\vh\equiv
  \frac{\partial {L}_{SD}}{\partial(\partial_0 \vh)}=
 -\int d^3x\left[ 4\vh~v_\psi~+2v_\vh\right]\equiv -[4\vh
 V_\psi+2V_\vh],
 $$
 while the zero Fourier harmonics of a canonical momentum of the spatial
 metric determinant  is
$$
 P_\psi\equiv-\int d^3x
 \frac{\partial{\cal L}_{SD}}{\partial(\partial_0 \log
 \overline{\psi})}=
 - \int d^3x{ \bar p}_{\psi}=\int d^3x
 \left[8\vh^2 v_\psi+4\vh~v_\vh\right]\equiv -2\vh[4\vh
 V_\psi+2V_\vh],
 $$
 where $V_\vh=\int d^3x ~v_\vh,V_\psi=\int d^3x ~v_\psi$. These two equations
 have no solutions, as
  the matrix  of the transition
from ``velocities'' to momenta
   has the zero  determinant.
 This means that the ``velocities''
  $[V_\vh,V_\psi]$ could not
  be expressed in terms of
  the canonical momenta
  $[P_\vh,P_{\psi}]$  and
 the Dirac Hamiltonian approach becomes a failure.
 To be consistent with identity
 (\ref{non1}) and to keep the number of variables of GR, we should
 impose the strong constraint
 \be
 V_\psi\equiv\int d^3 x {v_\psi}\equiv 0,
 \ee
 otherwise we shall have the double counting of the zero-Fourier
 harmonics of spatial metric determinant.

 What is double counting?
  A {\it``double counting''} is replacement of
  $L_1=(\dot x)^2/2$ by $L_2=(\dot x+\dot y)^2/2$.
 The second theory is not mathematically equivalent to the first.
 The test of this nonequivalence is the failure
 of the Hamiltonian approach to $L_2=(\dot x+\dot y)^2/2$.
Therefore,
  the replacement $L_1\to L_2 $ is nonsense in the context
  of the Hamiltonian approach.

 The next example   is Lifshitz's perturbation theory
 given by  Eq. (3.21) p. 217 in \cite{bard}
$$
ds^2= a^2(\eta)[(1+2\Phi)d\eta^2-(1-2\Psi)\gamma_{ij}dx^idx^j]
$$
 This
formula contains the double counting of the zero Fourier-harmonics
of the spatial metrics determinant presented by two variables: the
scale factor $a$ and $<\Psi>=\int d^3x\Psi(\eta,x_i)$ instead of
one.

 If we impose the strong constraint
$$
 <\Psi>=\int d^3x\Psi(\eta,x_i)\equiv 0;~~~
 <P_\Psi>=\int d^3x \partial L_{(2)}/\partial
 \dot\Psi(\eta,x_i)\equiv 0,
$$
 in order to remove the ``double counting'',
 we  shall return back to the Einstein theory (\ref{gr}).
 In  the Einstein theory, instead of
 the equations of $\Psi$ and $\Phi$ (4.15) on p. 220 in
 \cite{bard}
 \begin{align}
 \label{lif} -3{\cal H}({\cal H}\Phi+\Psi')+\triangle\Psi=&
 4\pi G\delta T^0_0 \\
 3[(2{\cal H}'+{\cal H}^2)\Phi+{\cal H}\Phi'+\Psi''+2{\cal
 H}\Psi']+\triangle(\Phi-\Psi)=&-4\pi G~\delta T^i_i,\nonumber
 \end{align}
 where $4\pi G=3/{(2\varphi^2)}$,  ${\cal H}=a'/a$, and
 $\triangle=\partial_i^2$,
 we shall obtain in Section 4.10 the
equations
 \begin{align}\label{0lif}
\triangle\Psi=& ~4\pi G~\delta T^0_0\\
\triangle(\Phi-\Psi)=&-4\pi G~\delta T^i_i.\label{10lif}
 \end{align}
 These Einstein equations will  not contain
 the time derivatives
 that are responsible for the ``primordial power spectrum'' in
  the inflationary model \cite{bard}.

\subsection{\label{hr} Hamiltonian approach }

 The  Hamiltonian action of GR  (\ref{1gr}),  (\ref{sv11}) in the field space
 $[\vh|{F}]$ takes the form \cite{pp,bpp}
 \be\label{12ha1}
 S=\int dx^0\left[-P_{\vh}\partial_0\vh+
 N_0\frac{P^2_\vh}{4V_0}+\int d^3x
 \left(\sum\limits_{{F}
 } P_{F}\partial_0\overline{F}
 +{\cal C}-N_d {\cal H}_t\right)\right],
 \ee
 where $P_{ F}$ is the set of the field momenta
 $\overline{p_{\psi}}, p^i_{{(a)}},p_f$;
 the sum of constraints
 ${\cal C}=N_{(a)} {T^0_{(a)}}_t +C_0\overline{p_\psi}+
C_{(a)}\partial_k{\bf e}^k_{(a)}$ contains  the weak Dirac
constraints
 of transversality  $\partial_i {\bf e}^{i}_{(a)}\simeq 0$
and the minimal space-like surface \cite{dir}
$\overline{p_{\psi}}={8\vh^2}v_{\psi}\simeq 0$
  with the Lagrangian multipliers $C_0,~C_{(a)}$;
 \be\label{2ha4}
 {\cal H}_t=\frac{1}{\vh^2}\left[{6p_{(ab)}p_{(ab)}}
 -{16}\overline{p_{\psi}}^2\right]
 +\frac{\varphi^2\overline{\psi}^{7}}{6}~
 \left[{}^{(3)}R({\bf e})\overline{\psi}+
 8\triangle\overline{\psi}\right]+\overline{\psi}^{4}T^0_{0({\rm M})}
 \ee
 is the Dirac Hamiltonian density \cite{dir} in terms of
 $p_{(ab)}=\left[p^i_{(a)}{\bf e}_{(b)i}+p^i_{(b)}{\bf
 e}_{(a)i}\right]/2$,
 \be\label{t0k}
 {T^0_{(a)}}_t=-\overline{p_{\psi}}\partial_{(a)} \overline{\psi}
 +\frac{1}{6}\partial_{(a)}
 (\overline{p_{\psi}}\overline{\psi})-\partial_{(b)}p_{(ba)}
 -p_{(bc)}{\bf e}_{(c)i}(\partial_{(b)}{\bf e}^i_{(a)}-
 \partial_{(a)}{\bf e}^i_{(b)})+ T^0_{{(a)}({\rm M})}~,
 \ee
 and $T^0_{0({\rm M})}$, $T^0_{k({\rm M})}$ are
 components of the energy--momentum tensor
 in terms of the York conformal fields $F_{\rm Y}^{(n)}
 =(a\psi^2)^nF^{(n)}$ \cite{Y}.

 The gauge-invariant lapse function $\overline{N_d}/N_0={\cal N}$ and
   the spatial metric determinant $\overline{\psi}$ can be determined by
   their equations for both the zero Fourier harmonic
   $\langle F\rangle$ and the nonzero ones $\overline{F}=F-\langle
   F\rangle$:
 \be\label{f0}
 ~~~~~~~\left\langle \overline{N_d}\frac{\delta S[\vh]}{\delta  \overline{N_d}
 }\right\rangle=0~~
 |\!\!\models\!\!\!\!\!\!
 \Longrightarrow\vh'^2=\rho_t,~~~~~~~~~~~~~~~~~~~~~~~~
 \overline{N_d\frac{\delta S[\vh]}{\delta  N_d}}=0~~
 |\!\!\models\!\!\!\!\!\!\Longrightarrow
 \frac{\rho_t}{\cal N}={\cal N}{\cal H}_t,
 \ee
 \be
 \left\langle
 \overline{\psi}\frac{\delta S[\vh]}{2\delta \overline{\psi}}\right\rangle
 =0~~
 |\!\!\models\!\!\!\!\!\!
 \Longrightarrow(\vh^2)''=3(\rho_t-p_t),~~~~~~~~~~~~
 \overline{\overline{\psi}\frac{\delta S[\vh]}{2\delta \overline{\psi}}}
 =0~~
 |\!\!\models\!\!\!\!\!\!
 \Longrightarrow~ \overline{\hat{\bf A}{\cal N}}=0
 \label{4f2},
 \ee
 where  $\varphi'\equiv{d\varphi}/{d\zeta}$,
 $\rho_t\equiv\langle{\cal N}{\cal H}_t\rangle$ and
 $p_t=\langle{\cal N}\overline{\psi}^{4}{T^k_k}_t\rangle/3$
 are the energy density and pressure of all fields,
 respectively;  and
 $\hat {\bf A}$ is a differential operator:
 $$
 \hat {\bf A} {\cal N}\equiv\frac{2\varphi^2}{3}\!
 \left[({}^{(3)}R({\bf e})\overline{\psi}^8\!+\!
 8\overline{\psi}^{7}\triangle\overline{\psi}) {\cal N}\!+\!
  \partial_j[\overline{\psi}^2\partial^j (\!\overline{\psi}^6
 {\cal N})]\right]\!+\!\overline{\psi}^{4}
 [3T^0_{0({\rm M})}\!-\!T^k_{k({\rm M})}]{\cal N}.
 $$
  Equations (\ref{f0}) and (\ref{4f2}) show us  that their zero harmonics
  coincide with the conformal version of the Friedmann
  equations with the scale factor $a=\vh/\vh_0$ \cite{pp}.
  In this case the Hamiltonian cosmological perturbation theory
   does not require
 its convergence to be proved because the perturbations are in a
 different class of functions (with nonzero Fourier harmonics) than
 the cosmological dynamics described by the equations in the zero
 harmonic sector.

  \subsection{\label{1hr} Dirac's Hamiltonian reduction }
 The energy constraints (\ref{f0}) have solutions
 \be\label{p}
 P_{\vh(\pm)}=\pm 2V_0\vh'=\pm 2V_0{\langle \sqrt{\cal
 H}_t\rangle},~~~~~~~{\cal N}={{\langle \sqrt{\cal H}_t\rangle}}/{\sqrt{{\cal
 H}_t}}
  \ee
 expressing the reduced Hamiltonian $P_{\vh(\pm)}$ and lapse
 function ${\cal N}$
 through the field energy density ${\cal H}_t$ given by Eq.
 (\ref{2ha4}).
 If we substitute these solutions into the action (\ref{12ha1}) and
 solve the first equation of $P_{\vh}$ with respect to the time $\zeta$,
  we obtain
    the {\it``reduced action``} and {\it``interval''} (\ref{ght}):
 \bea\label{2ha2} S_{\pm}[\varphi|\varphi_0]|_{\rm
constraint} &=&
 \int\limits_{\vh}^{\vh_0}d\vh \left\{\int d^3x
 \left[\sum\limits_{  F}P_{  F}\partial_\vh F
 +\bar{\cal C}\mp2\sqrt{{\cal H}_t}\right]\right\},\\\label{zi1}
\zeta_{\pm}[\varphi|\varphi_0]|_{\rm constraint} &=& \pm
\int\limits_{\vh}^{\vh_0}\frac{d\vh}{\langle\sqrt{{\cal
 H}_t}\rangle},
\eea
 where   $\bar{\cal C}={\cal
 C}/\partial_0\vh$ and
 $\vh_0$ is the present-day datum.
  Action (\ref{2ha2}) determines the
   evolution of fields directly in terms of the redshift parameter
  connected with the scale factor $\vh$ by the relation $\vh=\vh_0/(1+z)$
 and interval (\ref{zi1}) gives the Friedmann-like cosmic
 evolution. The Dirac constraint $\overline{p_\psi}=0$ in Eq. (\ref{2ha4})
  leads to
 the Hermitian reduced Hamiltonian.
The Dirac Hamiltonian {\it``reduction''}
 of the GR action  (\ref{h1}) onto its values  (\ref{2ha2}) obtained
  by the explicit resolution of the energy constraints
    determines
 main concepts of the primary quantization and secondary one, whereas
  the {\it``reduced interval''}   (\ref{zi1})
  gives  us the opportunity to
 clear up  the status of the Hubble evolution
 in the Hamiltonian theory.
 As we have seen in Section 3, the corresponding quantum theory
  describing the cosmological
  creation of a universe from the Bogoliubov stable
 vacuum  explains the absolute beginning of time and  removes
 a cosmological singularity. In other words, there are
 reasons to treat the evolution of the cosmological scale
 factor as a quantum collective motion of a system of all fields
 as the whole of the type of  the phenomenon of
 {\it``superfluidity''} \cite{Lon,Lan,B}.

 \subsection{\label{cpt} Hamiltonian cosmological perturbation theory}

 The Hamiltonian cosmological perturbation theory \cite{242} is defined using
 the decomposition of the forms (\ref{adm})
\be\label{ncpt} \omega_{(0)}=a(1+ \Phi)d\eta,~~~~
\omega_{(a)}=a(1-\Psi)
(dx_{(a)}+h^{(TT)}_{(a)i}dx^i+N_{(a)}d\eta),
 \ee
  where $a=\vh/\vh_0$ is the cosmological
 factor,
$N_{(a)}=\partial_{(a)}\sigma +N^{(T)}_{(a)}$. We take into
account the Dirac constraints of transversality $\partial_i
h^{(TT)}_{(a)i}=0$ and minimal surface
$\overline{p_\psi}=8\vh^2v_\psi\simeq 0$, where $v_\psi$ is given
by Eq. (\ref{pi1}). The latter defines the longitudinal shift
vector (\ref{ncpt}):
 \be\label{min10}
N^{||}_{(a)}=\partial_{(a)}\sigma;~~~~~~~~~~~~~
\triangle\sigma=-\frac{3}{4}\Psi'.\ee
  Therefore,
 there are six independent components: two scalars $\Phi$ and
$\Psi$, two vector ones $N^{(T)}_{(a)}$,  and two tensor ones
$h^{(TT)}_{(a)i}$.

The cosmological perturbations of the
 metric components are defined in the class
of functions
 with the nonzero Fourier harmonics
$\widetilde{\Phi}(k)=\int d^3x \Phi(x)e^{ikx}$
  (satisfying the strong constraint
 $\int d^3x \Phi(x)\equiv 0$).

 In the approximation
 $$\vh^2k^2\gg\rho_s=\langle T^0_{0\rm (M)}\rangle\gg
 \delta T_{0}^0=(T_{0\rm (M)}^0-\langle T^0_{0\rm (M)}\rangle);
 $$
 $$
 \vh^2k^2\gg 3p_s=\langle T^k_{k\rm (M)}\rangle \gg\delta T^k_{k}
 =(T_{k\rm (M)}^k-\langle T^k_{k\rm (M)}\rangle),
 $$
 six equations of
 the theory
 for the scalar, vector,  and tensor components   take
 form
  \begin{align}
 \label{107}
 \triangle{\Psi}&=
 4\pi G~\delta{T}^0_{0},\\
 \triangle
{\Phi}
&=4\pi G~(\delta{T}^0_{0}-\delta{T}^j_{j}),\label{108}\\
\frac{1}{4}\triangle N^{T}_{j}&=-4\pi G\delta{T}^{0\:(T)}_j,\\
\frac{1}{8}\left[-\triangle
 h^{(TT)}_{ij}+
 \frac{(\varphi^2{h^{TT}_{ij}}')
 {\vphantom{h^{TT}_{ij}}}'}{\varphi^2}\right]&=
 4\pi G~\delta{T}^{TT}_{ij},
 \end{align}
 where
 $ \partial_i \delta T^{(TT)}_{ij}=0,~~ \delta T^{(TT)}_{ii}=0,
 ~~ \partial_j\delta T_j^{0\:(T)}=0$, and  $4\pi G={3}/({2\vh^2})$ is
 running Newton coupling ``constant''.

 Eqs. (\ref{107}) and (\ref{108}) differ from the standard
cosmological perturbation theory \cite{Lif,bard} by the absence of
the time derivative of deviations of the linear spatial metric
 determinant $\Psi$ and lapse function  $\Phi$ (see Eqs. (4.15) in \cite{bard}).
 These time derivatives,  in the Lifshitz
 perturbation theory \cite{Lif,bard}, go from the kinetic term ${\bf K}$
 and from the zero Fourier
 harmonics of the trace of second
 form. In the Hamiltonian perturbation theory (with the Dirac minimal surface
 $\overline{p_\psi}=0$ that leads
 to the Hermitian Hamiltonian in the field space of events
 $[\vh|F]$),
 the kinetic term ${\bf K}$ in action (\ref{sv11})
contributes only to the
 next orders of the perturbation theory, while the ``quasi--surface'' term
  $(\int d^3 x {\bf S})$    was removed from action (\ref{sv11})
  in order to formulate the
 Hamiltonian approach to GR
  without  the double counting
 of the  scale factor velocity.
   The potential term ${\bf P}$ leads only to the spatial
 derivatives.

  The solutions of
 (\ref{107}) and (\ref{108}) take the form of
standard classical solutions with the Newton
 gravitational
 constant
$G= {3}/{8\pi\varphi^2}$ (see Eqs. (\ref{0lif}) and
(\ref{10lif})):
 \be\label{dgg}
 \widetilde{\Psi}=-\frac{4\pi
 G}{k^2}\delta \widetilde{T}^{0}_{0};~~~~~~~~~~~
 \widetilde{\Phi}
 =-\frac{4\pi G}{k^2}\left[\delta \widetilde{T}^{0}_{0}
 -\delta \widetilde{T}^{k}_{k}\right].
 \ee
 The minimal surface $\overline{p_\psi}=0$ (\ref{min10})
 gives the shift of the coordinate
  origin in the process of evolution, in particular, in the case of
   a point source
  $\delta T^0_0=M_J[\delta^3(x-y_J)-1/V_0]$,
  we got the shift vector:
 \be\label{ni}
 N^i=-\frac{3(GM_J)'}{4}
 \frac{(x-y_J)^i}{|x-y_J|}.
  \ee
Interval (\ref{ncpt})
  determines an equation for the photon momenta
  \be \label{12ni}
  p_{\mu}p_{\nu}g^{\mu\nu}\simeq (p_0+N^ip^i)^2(1-2\Phi)
  -p^2_j(1+2\Psi)=0,
 \ee
 from which we obtain a photon energy
  \be
 p_0\simeq-N^ip^i+[1+({{\Phi}+{\Psi}})]|p |;~~
 |p |=\sqrt{p^2_i}.
 \ee
 This formula shows us
 the relative magnitude of  spatial fluctuations of a photon energy
 in terms of  the metric components  $
 {p_0-|p|}/{|p|}=-[N^in^i+({{\Phi}+{\Psi}})]
 $, $n^i={p_i}/{|p|}$.
 The appearance of the spatial anisotropy (\ref{ni})
  in the flow of the photon energy  is
 the consequence of the minimal surface  $\overline{p_\psi}=0$, and
 this anisotropy (\ref{12ni})   can be taken account
 in order to describe the spectrum of CMB temperature fluctuation.

 \subsection{\label{ept} Bogoliubov's quantum Gravity}

 Quantum theory for GR
 is based, on the one hand, on the existence of
 the Hilbert geometric formulation \cite{H}  of Special
 Relativity considered in Section 2 as the simplest model of GR
 and, on the other hand, on  the contemporary quantum field theory
  appearing as a result of resolution of the energy constraint
  and its primary and secondary quantizations.

  Resolution of
  the energy constraint leads to
 the  {\it``Hamiltonian reduction''} of the Hilbert geometric action
 (\ref{2ha2}), and it
    determines the  {\it``reduced energy''}
 $E_\vh=2V_0{\langle\sqrt{{\cal H}_t}\rangle}$
 as the central
 concept of the primary
 quantization $\hat P_\vh\Psi=-i\partial_\vh\Psi$.
 The primary
 quantization
  converts the energy constraint $P^2_\vh-E_\vh^2=0$ into
 the WDW equation: $\partial^2_\vh\Psi+E_\vh^2\Psi=0$ \cite{WDW}.

 The next step
 is the secondary quantization, where $\Psi=[A^++A^-]/({\sqrt{2E_\vh}})$
  is considered as a ``quantum field'' with {\it ``one-universe energy''}
 $E_\vh$. We have seen in Section 3 that the Bogoliubov transformation
 $A^+=\alpha
 B^+\!+\!\beta^*B^-$ \cite{B} of
 operators of {\it ``universe''} $A^+,A^-$
 to ones of {\it``quasiuniverse''} $B^+,B^-$
   allows as  to postulate the
{\it vacuum} as a stable state of the minimal energy
$B^-|0>_{U}=0$,
  to find conserved numbers of {\it ``quasiuniverses''} ${\cal
N}_B=(B^+B^-)$
 and calculate a distribution function of creation of the
 {\it ``universe''} $N_{\rm U}(\vh)={}_{\rm U}<A^+A^-
  >_{\rm U}\equiv |\beta|^2$ and the ``rotation function''
 $R_{\rm U}=i(\alpha^*\beta-\alpha\beta^*)$
satisfying equations \cite{ps1}: \be\label{10usv3}
 \frac{dN_{\rm U}}{d\vh}=-\frac{\partial_{\varphi}E_\vh}{4E_\vh^2}
 \frac{dR_{\rm U}}{d\vh},~~~~~~
 \frac{dR_{\rm U}}{d\vh}=-{2E_\vh}
 \sqrt{4N_{\rm U}(N_{\rm U}+1)-R_{\rm U}^2}
 \ee
 with the initial data $N_{\rm U}(\vh=\vh_I)=R_{\rm
 U}(\vh=\vh_I)=0$.

 The ``vacuum'' postulate  restricts the motion of the
 universe in the field space of events $[\vh|F]$: a universe moves
 forward $\vh>\vh_I$ for positive energy
 $P_\vh\geq 0$ (creation of a universe), and a universe
 moves backward $\vh<\vh_I$ for for negative $P_\vh\leq 0$
  (annihilation of a universe), where $\vh_I$ is the initial data.
This restriction leads to positive arrow of the ``interval''
(\ref{zi1}) $\zeta_{\pm}\geq 0$ and its absolute beginning
\cite{pp}.

\subsection{ Einstein's correspondence principle and relative units}

 Einstein's correspondence principle \cite{pp}
 as the low-energy
 expansion of the {\it``reduced action''} (\ref{2ha2}) over the
 field density ${\cal T}_{{\rm s}0}^0$
 $$d\vh 2\sqrt{{\cal H}_t}= d\vh
 2\sqrt{\rho_{0}(\vh)+{\cal T}_{{\rm s}0}^0}
 =
 d\vh
 \left[2\sqrt{\rho_0(\vh)}+
 {{\cal T}_{{\rm s}0}^0}/{\sqrt{\rho_0(\vh)}}\right]+...$$
 gives the sum:
 $S^{(+)}[\varphi_I|\varphi_0]|_{\rm
 constraint}= S^{(+)}_{\rm cosmic}+S^{(+)}_{\rm
 field}+\ldots$, where
 $S^{(+)}_{\rm cosmic}[\varphi_I|\varphi_0]= -
 2V_0\int\limits_{\vh_I}^{\vh_0}\!
 d\vh\!\sqrt{\rho_0(\vh)}$ is the reduced  cosmological action (\ref{ham-2}),
 and
 \be\label{12h5} S^{(+)}_{\rm field}=
 \int\limits_{\eta_I}^{\eta_0} d\eta\int d^3x
 \left[\sum\limits_{ F}P_{ F}\partial_\eta F
 +\bar{{\cal C}}-{\cal T}_{\rm s 0}^0\right]
 \ee
 is the standard field action
 in terms of the conformal time:
 $d\eta=d\vh/\sqrt{\rho_0(\vh)}$,
 in the conformal flat space--time with running masses
 $m(\eta)=a(\eta)m_0$ that describes the cosmological particle creation
 from vacuum \cite{ps1}.

 \begin{figure}[t]
\vspace{1cm}
 \begin{center}
 \includegraphics[width=0.65\textwidth,clip]{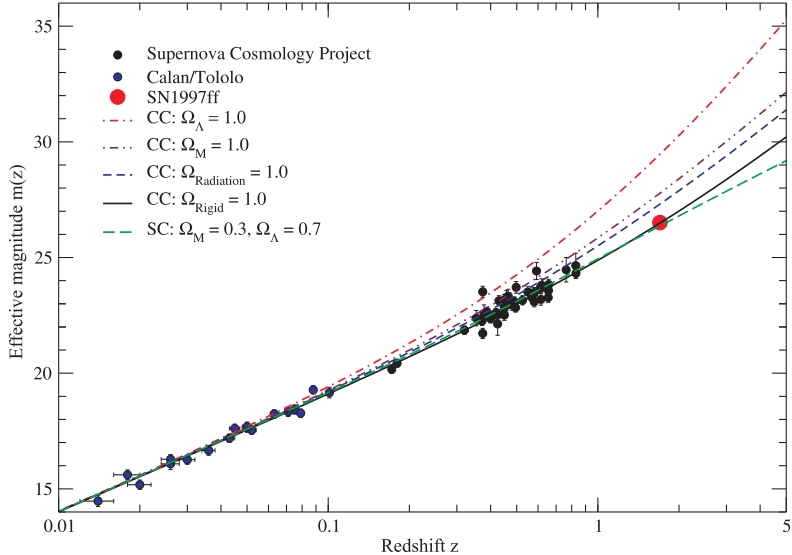}
\caption{ The Hubble diagram~\cite{039a,Danilo} in cases of the
{\it``absolute''} units of standard cosmology (SC)  and the
{\it``relative''} ones of conformal cosmology (CC).
 The points include  42 high-redshift Type Ia
 supernovae~\protect\cite{snov} and the reported
 farthest supernova SN1997ff~\protect\cite{SN}. The best
fit to these data  requires a cosmological constant
$\Omega_{\Lambda}=0.7$ $\Omega_{\rm Cold Dark Matter}=0.3$ in the
case of SC, whereas in CC
 these data are consistent with  the dominance of the rigid (stiff) state.
\label{fig1}}
\end{center}
\end{figure}

 This expansion shows us that the Hamiltonian approach
 identifies the ``conformal quantities''
  with the observable ones including the conformal time $d\eta$,
  instead of $dt=a(\eta)d\eta$, the coordinate
 distance $r$, instead of Friedmann one $R=a(\eta)r$, and the conformal
 temperature $T_c=Ta(\eta)$, instead of the standard one $T$. Therefore
 the correspondence principle  distinguishes the conformal cosmology (CC)
 \cite{Narlikar,039},
  instead of the standard cosmology (SC).
 In this case
 the
  red shift of the spectral lines of atoms on cosmic objects
 $$
\frac{E_{\rm emission}}{E_0}=\frac{m_{\rm atom}(\eta_0-r)}{m_{\rm
atom}(\eta_0)}\equiv\frac{\vh(\eta_0-r)}{\vh_0}=a(\eta_0-r)
=\frac{1}{1+z}
$$
is explained by the running masses $m=a(\eta)m_0$ in action
(\ref{12h5}).

The conformal observable distance  $r$ loses the factor $a$, in
comparison with the nonconformal one $R=ar$. Therefore, in the
 case of CC, the reduced interval (\ref{4conn})
  describing the redshift --
  coordinate-distance relation \cite{039} corresponds to a different
  equation
  of state than in the case of SC.
 The best fit to the data,  including
  Type Ia supernovae~\protect\cite{snov,SN},
 requires a cosmological constant $\Omega_{\Lambda}=0.7$,
$\Omega_{\rm Cold Dark Matter}=0.3$ in the case of the Friedmann
``absolute quantities`` of standard cosmology. In the case of
``conformal
 quantities'' in CC, the Supernova data \cite{snov,SN} are
consistent with the dominance of the stiff (rigid) state,
$\Omega_{\rm Rigid}\simeq 0.85 \pm 0.15$, $\Omega_{\rm
Matter}=0.15 \pm 0.15$ \cite{039,039a,Danilo}. If $\Omega_{\rm
Rigid}=1$, we have the square root dependence of the scale factor
on conformal time $a(\eta)=\sqrt{1+2H_0(\eta-\eta_0)}$. Just this
time dependence of the scale factor on
 the measurable time (here -- conformal one) is used for description of
 the primordial nucleosynthesis \cite{Danilo,three}.


 This stiff state is formed by a free scalar field
 when $E_\vh=2V_0\sqrt{\rho_0}=Q/\vh$. Just in this case there is an exact
solution of  Bogoliubov's equations (\ref{11cu})
 \be\label{cu}
 N_{\rm U}(\vh_0)=\frac{1}{4Q^2-1}
 \sin^2\left[\sqrt{Q^2-\frac{1}{4}}~~\ln\frac{\vh_0}{\vh_I}\right]\not
=0, \ee
 where the Planck mass $\vh_0=\vh_I\sqrt{1+2H_I\eta_0}$
 belongs to the present-day data $\eta=\eta_0$ and
 $\vh_I,H_I=\vh'_I/\vh_I=Q/(2V_0\vh_I^2)$ are the initial data.

\subsection{ Cosmological creation of matter}

 These initial data $\vh_I$ and $H_I$ are determined by the
 parameters of matter cosmologically created from the Bogoliubov
 vacuum  at the beginning of a universe $\eta\simeq 0$.

 The Standard
 Model (SM) density ${\cal T}_{{\rm s}0}^0$ in action (\ref{12h5})
  shows
 us that W-, Z- vector bosons have maximal probability of this
 cosmological creation
 due to their mass singularity~\cite{114:a}. One can introduce the notion of
 a particle in a universe if the Compton length of a particle
 defined by its inverse mass
 $M^{-1}_{\rm I}=(a_{\rm I} M_{\rm W})^{-1}$ is less than the
 universe horizon defined by the inverse Hubble parameter
 $H_{\rm I}^{-1}=a^2_{\rm I} (H_{0})^{-1}$ in the
 stiff state. Equating these quantities $M_{\rm I}=H_{\rm I}$
 one can estimate the initial data of the scale factor
 $a_{\rm I}^2=(H_0/M_{\rm W})^{2/3}=10^{29}$ and the primordial Hubble parameter
 $H_{\rm I}=10^{29}H_0\sim 1 {\rm mm}^{-1}\sim 3 K$.
 Just at this moment there is  an effect of intensive
  cosmological creation of the vector bosons described in \cite{114:a};
 in particular, the distribution functions of the longitudinal   vector bosons
demonstrate us a large contribution of relativistic momenta, as it
was shown in Fig.~2.
 Their conformal (i.e. observable) temperature $T_c$
 (appearing as  a consequence of
 collision and scattering of these bosons) can be estimated
from the equation in the kinetic theory for the time of
establishment of this temperature $ \eta^{-1}_{relaxation}\sim
n(T_c)\times \sigma \sim H $, where $n(T_c)\sim T_c^3$ and $\sigma
\sim 1/M^2$ is the cross-section. This kinetic equation and values
of the initial data $M_{\rm I} = H_{\rm I}$ give the temperature
of relativistic bosons
\be\label{t}
 T_c\sim (M_{\rm I}^2H_{\rm I})^{1/3}=(M_0^2H_0)^{1/3}\sim 3 K
\ee
as a conserved number of cosmic evolution compatible with the
Supernova data \cite{039,snov,SN}.
 We can see that
this  value is surprisingly close to the observed temperature of
the CMB radiation
 $ T_c=T_{\rm CMB}= 2.73~{\rm K}$.


\begin{figure}
  \centering
  \includegraphics{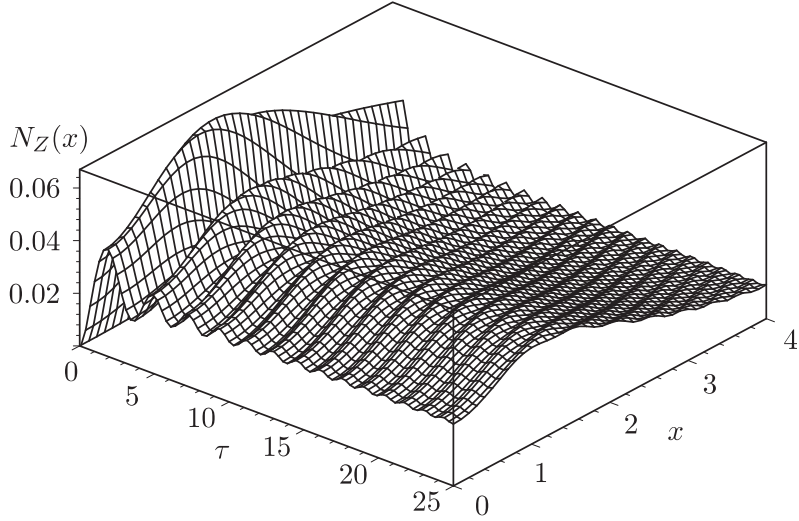}
  \caption{Longitudinal ($N_Z(x)$) components
of the boson distribution versus the dimensionless time $\tau=
2\eta H_I$ and the dimensionless momentum $x = q/M_I$ at the
initial data $M_I = H_I$ ($\gamma_v = 1$) \cite{114:a,vin}.}
\end{figure}

 The primordial mesons before
 their decays polarize the Dirac fermion vacuum
 (as the origin of axial anomaly \cite{riv,ilieva,gip,j})
 and give the
 baryon asymmetry frozen by the CP -- violation.
The
 value of the baryon--antibaryon asymmetry
of the universe following from this axial anomaly was estimated in
\cite{114:a} in terms of the coupling constant of the
superweak-interaction
 \be\label{a}
 n_b/n_\gamma\sim X_{CP}= 10^{-9}.
 \ee The boson
life-times     $\tau_W=2H_I\eta_W\simeq
\left(\frac{2}{\alpha_g}\right)^{2/3}\simeq 16,~ \tau_Z\sim
2^{2/3}\tau_W\sim 25
 $ determine the present-day visible
baryon density
\be\label{b}\Omega_b\sim\alpha_g=\alpha_{QED}/\sin^2\theta_W\sim0.03.\ee

 This baryon density as a final product of the  decay of bosons
 with momentum $q$ and
 energy $\omega(\eta)=(M^2(\eta)+q^2)^{1/2}$ oscillates as
 $\cos\left[2\int_0^{\eta}d\bar \eta \omega(\bar \eta) \right]$ \cite{114:a}.
 One can see  \cite{vin} that the number  of  density oscillations of the
 primordial bosons during their life-time for the momentum $q\sim
 M_I$  is of order of $20$, which is very close to the number of
 oscillations of the visible baryon matter density recently
 discovered in researches of large scale periodicity in redshift
 distribution \cite{a1,a2}
 \be\label{ls}
 [H_0\times128~{\rm MPc}]^{-1}\sim 20\div25\sim (\alpha_g)^{-1}.
 \ee
  The results (\ref{t}), (\ref{a}), (\ref{b}), (\ref{ls})
 testify to that all  visible matter can be a product of
 decays of primordial bosons with the oscillations forming
 a large-scale structure of the baryonic matter.

 The temperature history of the expanding universe
 copied in the ``conformal quantities'' looks like the
 history of evolution of masses of elementary particles in the cold
 universe with conformal temperature $T_c=a(\eta)T$
 of the cosmic microwave background.

\subsection{ The Dark Matter problem}

In the considered model, galaxies and their clusters are formed by
the Newton Hamiltonian with running masses
$E(\eta)={p^2}/{2m(\eta)}-{r_g(\eta)m(\eta)}/2r$, where the Newton
coupling ${r_g(\eta)m(\eta)}/{2}=r_g(\eta_0)m(\eta_0)/2$ is a
motion constant. One can see that the running masses lead to the
effect of the capture of an object by a gravitational central
field at the time when $E(\eta_{\rm capture})=0$. After the
capture the conformal size of the circle trajectories decreases as
${r(\eta)=R_0/a(\eta),~R_0=\rm const}$.

 The running masses change the orbital curvatures \cite{dm}
 \be\label{orb1}
 v_{\rm orbital}(R_0)=\sqrt{\frac{r_g}{2R_0}+\gamma(R_0 H)^2},
 \ee
 where
 $$
\gamma=2-\frac 32 \Omega_{\rm Matter}- 3 \Omega_\Lambda
 $$
 is determined by the equation of state. We can see that in the
 case of
 the stiff state of the conformal cosmology, when
 $\Omega_{\rm Matter}=\Omega_\Lambda=0$ and $\gamma=2$,
  the cosmological evolution   plays the role of the Dark Matter,
   and it can explain the  deficit of the
luminous matter $M/M_{L}\sim 10^2$, where $M_L$ stands for the
mass of luminous matter, in superclusters with a mass $M\geq
10^{15}M_\odot$, $R\gtrsim 5 {\rm Mpc}$ \cite{dm}, where the
Newton velocity becomes less than the cosmic one.
 In the case of standard
cosmology: $\Omega_{\rm Matter}=0.3$, $\Omega_\Lambda=0.7$, the
last term is negative  $\gamma=-1/2$. Thus, the standard cosmology
requires one more Dark Matter in contrast to the conformal
cosmology \cite{039}.

  However the cosmological modification of the Newton dynamics
  given by Eq. (\ref{orb1}) is not sufficient to explain the
  constant orbital velocities in spiral galaxies\footnote{The cylindric symmetry of matter sources in
  spiral galaxies  points out  that their gravitational potential
    can be the two-dimensional Newton one
    $\triangle_{(2)} \Phi=(1/2l)\delta^{2} (x)$ with the
 length of an axis $2l$ \cite{dm}.
 This potential leads to the corresponding orbital velocity
  $v^2_{\rm orbital}(R_0)={r_g/\left(2\sqrt{l^2+R_0^2}\right)}$,
  and it
 can explain the constant rotational curves for spiral galaxies
  in the region $R_0 \lesssim l$ in the case of $\Omega_b\sim 5\%$.}.


\section{Conclusion}

 We have seen that GR could pass along the pathway of quantum field theory
 through the Dirac reduction and the Bogoliubov transformation
 (see the table on p. \pageref{gd}), in order to describe
 the cosmological creation of universes (\ref{10usv3}). This pathway
 includes the zero mode of a general resolution of
  constraints in the class of functions of gauge transformations
  (as the global homogeneous excitation
  of the type of
   Landau superfluid liquid), the WDW unique wave function as
  London's attribute of superfluidity,
  the postulate of the quantum Bogoliubov vacuum leading to
   the absolute beginning of geometric time, the Einstein correspondence
   principle identifying the conformal quantities with the ``measurable''  ones,
   and the uncertainty  principle for establishing the point of
   the beginning of the cosmological creation of the primordial W-, Z- bosons
  from vacuum due to their mass singularity.

 The Hamiltonian approach revealed the  double counting
 of the cosmological scale factor in
   the standard Lifshitz
 perturbation theory. It  means that this standard
 perturbation theory
  does not coincide with the Einstein theory.
Avoiding this  double counting, in order  to return back to GR,
 we have obtained new Hamiltonian  equations. These  equations do not contain
 the time derivatives
 that are responsible for the ``primordial power spectrum'' in
 the inflationary model \cite{bard}.
 However, Dirac's Hamiltonian approach to GR gives  us
 another  possibility  to explain this ``spectrum'' and
  other topical problems of cosmology
  by  the cosmological creation of the primordial W-, Z- bosons
  from vacuum when
 their Compton length coincides with the universe horizon.

 The equations describing the longitudinal
 vector bosons
 in SM, in this case, are close to the equations
 of the inflationary model used for
 description of the ``power primordial spectrum'' of the CMB radiation.
 We listed the set of theoretical and observational
 arguments in favor of that the CMB radiation can be
 a final product of  primordial vector W-, Z- bosons cosmologically created
 from Bogoliubov vacuum.

 This pathway of quantization points out that in GR and SM there is
    a new
principle of relativity - a relativity of  units of measurements.
It means that equations of motion  do not depend not only on the
data but also on the units of measurement of these data. In
context of this principle of relativity one can propose that the
Higgs potential is not necessarily \cite{PR}.

\vspace{1cm}

{\bf Acknowledgments}

\medskip

 The authors are grateful to   B.M. Barbashov, A.A. Gusev, A.V. Efremov, P. Flin,
 E.A. Kuraev, L.N. Lipatov, D. Mladenov,  G.M. Vereshkov, S.I. Vinitsky,
  and
 A.F. Zakharov for interesting and critical
discussions.

\end{document}